\documentclass[twocolumn,english,floatfix,prl]{revtex4}
\usepackage[T1]{fontenc}
\usepackage[latin1]{inputenc}
\usepackage{graphicx}

\makeatletter

\usepackage{slashbox}
\usepackage{hhline}
\usepackage{psfrag}

\usepackage{babel}
\makeatother
\begin{document}

\title{Are Domain Walls in 2D Spin Glasses described by Stochastic
Loewner Evolutions?}

\author{Denis Bernard$^{1}$, Pierre Le Doussal$^{1}$, A. Alan
Middleton$^{2}$ }

\affiliation{$^{1}$CNRS-Laboratoire de Physique Théorique de l'Ecole
Normale Supérieure, 24 rue Lhomond, 75005 Paris, France.}

\affiliation{$^{2}$Department of Physics, Syracuse University, Syracuse, NY 13244,
USA}

\begin{abstract}
Domain walls for spin glasses are believed to be scale
invariant; a stronger symmetry, conformal
invariance, has the potential to hold. The statistics of
zero-temperature Ising spin glass domain walls in two dimensions
are used to test the hypothesis that these domain walls are
described by a Schramm-Loewner evolution $\mathrm{SLE}_\kappa$.
Multiple tests
are consistent with $\mathrm{SLE}_\kappa$, where $\kappa=2.32\pm0.08$.  Both
conformal invariance and the domain Markov property are tested.
The latter does not hold in small systems, but detailed
numerical evidence suggests that it holds in the continuum
limit.
\end{abstract}

\maketitle The geometrical characterization of
physical objects is central to much of our understanding of their
energetics and dynamics.  The relevant geometries can be as simple as
points or gently curved surfaces. Many
objects are not well-described by an integer dimension, but have a
scale-dependent measure that can be represented by a fractal
dimension. For example, continuous
phase transitions in homogeneous systems have nonanalytic behavior
consistent with
fractal dimensions for the surfaces that separate phases.
Evidence for fractal domain walls
is seen in scattering experiments and numerical simulations. In models
of glassy systems with quenched disorder (frozen-in random fields),
analytic work and numerical simulations indicate that domain
walls can be scale-invariant and fractal at low temperatures.
In two-dimensional homogeneous systems,
the additional symmetry of conformal invariance often applies and
yields detailed predictions for critical
exponents, the effects of boundary conditions, and a
background for physical explanations.

The conjunction of conformal invariance with the presence of a
domain Markov property (DMP) in statistical mechanics models has led
to an even more complete - and in several cases mathematically
rigorous - description of fractal curves such as loop-erased
self-avoiding walks, percolation hulls, and domain walls at phase
transitions in two dimensions in the scaling (i.e., continuum)
limit \cite{CardyReview,BauerBernard}.
Schramm showed that when both properties are present the
probability measure on these curves is described by a Schramm-Loewner
evolution $\mathrm{SLE}_\kappa$ \cite{schramm}.  Random sequences of simple
conformal maps can be used to generate the fractal curves with the
correct measure, if the real-valued driving function that underlies
the maps is a Brownian motion. The diffusion coefficient $\kappa$ of
the Brownian motion uniquely parameterizes the process and is related
to the fractal dimension of the curve via $d_f=1+\kappa/8$. This deep
connection has led to very precise characterization of these curves
for pure systems such as $q$-states Potts model, $O(n)$ models and
percolation.

An outstanding question is whether SLE can be applied to other
systems. Numerical evidence has been presented, for example, that SLE
describes certain isolines in 2D turbulence \cite{BernardFalkovichEtc}.
The broad question of whether and how conformal invariance, a
necessary condition for SLE, applies to disordered systems is still
very much open. Attempts to extend the apparatus of conformal field
theory to systems with quenched disorder, a notably difficult subject
\cite{cardy}, have suggested some numerical tests, such
as finite size scaling \cite{jacobsen}. A positive result was obtained
recently for surface wavefunction multifractality at the 2D
localization transition with spin-orbit symmetry \cite{gruzberg}. Most
importantly for our work, Amoruso, Hartmann, Hastings and Moore
\cite{amoruso} have suggested that domain walls in the 2D spin
glass have conformal invariance.

Here we directly investigate whether the domain wall statistics
converges to an $\mathrm{SLE}_\kappa$. We apply several tests.
We examine the
winding of domain wall around a cylinder, as well as
the angular distribution of
the curves and the dipolar SLE hitting probability. We use
an iterated slit map (discretized inverse Loewner evolution) to
determine the driving function and test whether it converges to
Brownian motion. We directly test the DMP by comparing precise domain
wall statistics in ``whole'' and ``cut'' domains.  To determine the
significance of these tests, we carry out the same analyses for the
loop-erased random walk (LERW), which has
$\mathrm{SLE}_2$ as a scaling limit, and for paths on minimal spanning
trees (MST), which are not conformally invariant \cite{Wilson}.  We
find that, for one choice of boundary conditions (BCs), the spin glass
domain wall passes all tests with a consistent value of $\kappa$.

We study the domain walls in a 2D Ising spin glass with Gaussian
disorder.  We use the Edwards-Anderson Hamiltonian ${\cal
H}=-\sum_{\langle ij\rangle}J_{ij}s_{i}s_{j}$, where $s_{i}=\pm 1$
and the $J_{ij}$ are each chosen from a
Gaussian distribution with zero mean. The glass
transition is at $T=0$; we study the minimum energy states at $T=0$
using an exact optimization algorithm \cite{Barahona} and sample
over disorder realizations. There are two
ground states, connected by a global spin flip, in any finite sample.
In the scaling picture based on domain walls and droplets, introduced
by McMillan \cite{mcmillan}, Bray and Moore \cite{BrayMoore}, and
Fisher and Huse \cite{fisherhuse}, there are two ground states in the
thermodynamic limit \cite{middleton,youngpalassini}.
Domain walls (DWs)
separate these two ground states.
The domain wall energy scales as $E_{{\rm
DW}}\sim L^{\theta}$ \cite{mcmillan,BrayMoore,fisherhuse} for DWs
defined at scale $L$, with
$\theta=-0.28(1)$.

\begin{figure}
\includegraphics[%
  width=1.0\columnwidth]{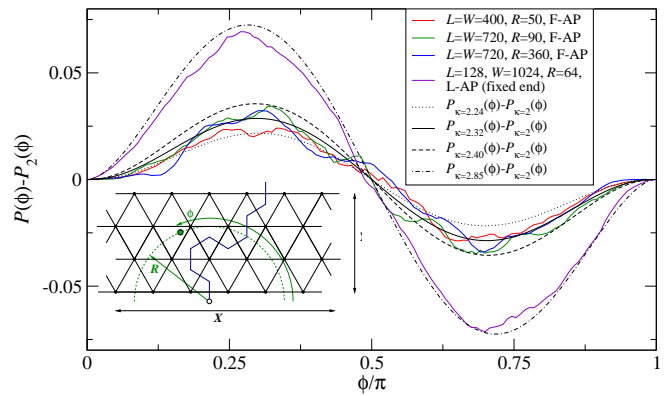}

\caption{\label{cap:angle} Plots of
$P(\phi,R)-P_{2}(\phi)$, where $P(\phi,R)$ is the probability for the
domain wall to pass to the left of a point with polar coordinates
$(R,\phi)$ (see inset). The magnitude of statistical errors (not shown) is
consistent with the apparent fluctuations of the data lines.  The
predicted $P_{2}(\phi)=[\phi-\frac{1}{2}\sin(2\phi)]/\pi$ for
$\kappa=2$ is subtracted to display small variations clearly.  The
data from F-PA paths agrees with SLE predictions for $\kappa$ in the
range $2.24<\kappa<2.40$, while L-PA paths give
$\kappa\approx2.85(10)$.  Inset: A domain wall of length
$S=9$ in a sample with $L=4$ rows and $W=6$ columns.}
\end{figure}

We work on a triangular lattice that has $W$ spins
in each of $L$ rows, as sketched in Fig.~\ref{cap:angle}.  Our
samples are cylindrical, with periodic rows.  One can uniquely
describe ground state pairs by the bond satisfactions
$\sigma_{ij}={\rm sgn}(J_{ij}s_{i}s_{j})$, where
$\Pi_{\Delta}\sigma_{ij}={\rm sgn}(\Pi_{\Delta}J_{ij})$ for any
elementary triangle $\Delta$. 
Periodic BCs result from fixing
$\Pi_{(ij)\in p_{\alpha}}\sigma_{ij}=\Pi_{(ij)\in p_{\alpha}}{\rm
sgn}(J_{ij})$, where $p_{\alpha}$ are the sets of boundary bonds on
the upper ($\alpha=1$) and lower ($\alpha=2$) edges
BCs.  Imposing $\Pi_{(ij)\in p_{\alpha}}\sigma_{ij}=-\Pi_{(ij)\in
p_{\alpha}}{\rm sgn}(J_{ij})$ gives antiperiodic BCs (equivalent to a
change in the sign of the $J_{ij}$ along a column of bonds). Comparing
ground states for periodic and antiperiodic BCs gives a domain wall: a
simple path on the dual lattice that crosses bonds whose
satisfaction differs between the two BCs.  A domain wall is the
minimizer of the cost function $2\sum_{(ij) \in
\gamma}F_{ij}$, with $F_{ij} = J_{ij} s_i^0 s_j^0$ and $s_i^0$
the spins in the periodic ground state, over
open paths $\gamma$ 
from the bottom to the top of the cylinder.  In the ground state, the
cost of any closed loop is positive \cite{footnote1}.

We refer to the domain wall found using this particular
periodic-antiperiodic BC comparison as {}``floating'' (F-PA), as the
endpoints of the domain wall are not fixed. We also consider a
periodic-antiperiodic BC change where the domain wall at one end is
locally constrained to a single chosen bond on the lower boundary (L-PA),
i.e., a {\em given} $\sigma_{ij}$ is reversed on the lower boundary.

The choice of $L$ and $W$ give the cylinder shape, with the
circumference given by $X=W$ and the length by $Y=(\sqrt{3}/2)L$.  We
find that the averages converge for $W\ge 4L$. The results
for the first quarter of the path for $W=L$ agree with those for $W\gg L$,
within our accuracy.  For
comparison, we also study LERW curves and paths between two points in
the MST, both on honeycomb lattices; LERW curves have
dimension $d_{f}=5/4$ in the continuum limit and MST
paths appear to have a fractal dimension $d_{f}=1.217(3)$ \cite{MST}.

We estimate $d_f$ for the domain wall by computing the mean total path
length $\overline{S}(L)$ of the domain wall, comparing with
$\overline{S}(L)\sim L^{d_f}$, the overline indicating averages over
$\approx 10^{4}$ samples at sizes up $5\times10^{5}$ spins, and also by
computing the sample
averaged distance from the origin
as a function of partial path length.  We find
$d_{f}=1.28(1)$, in
statistical agreement with previous work \cite{SG2D},
{\em for both F-PA and L-PA BC's}.

We test conformal invariance and consistency with the SLE description
by measuring the winding of the F-PA domain wall around a long
cylinders with $Y\gg X$.  The prediction from SLE is that the variance
of the transverse displacement $x$ of the end point from the starting
location is $\langle x^{2}\rangle=\frac{4}{\pi}(d_{f}-1)XY$.  We
studied cylinders with $8\le W \le 32$ and up to $L=800$ for at
least $10^{4}$ samples at each of at least eight values of $L$.  Our
data is consistent with $\frac{\pi}{4}\langle x^{2}\rangle$
linear in $XY$, with a coefficient of $0.27\pm0.01$ for $L>4W$, in
agreement with conformal invariance, {\em again for both F-PA and L-PA BC's}.

A powerful result \cite{schramm2} from SLE is a prediction of the
probability that a curve generated by SLE will pass to the left of a
given point at polar coordinates $(R,\phi)$
(see Fig.\ \ref{cap:angle} for notation).  Given scale invariance,
the probability that the curve passes to the left of $(R,\phi)$
depends only on $\phi$, and the theory of SLE can be used to predict
\cite{schramm2}
\begin{equation}
P_{\kappa}(\phi)=\frac{1}{2}+\frac{\Gamma\left(\frac{4}{\kappa}\right)}{\sqrt{\pi}
\Gamma\left(\frac{8-\kappa}{2\kappa}\right)}\tan(\phi)F_{12}
\left(\frac{1}{2};\frac{4}{\kappa},\frac{3}{2};-\tan^{2}(\phi)\right)\,,\label{eq:angle}
\end{equation}
where $F_{12}$ is the hypergeometric function and $\kappa$ is the
diffusion parameter from continuum SLE. Our results
(Fig.~\ref{cap:angle}) for $P(\phi)$ depend on the choice of BC.
For F-PA BCs the measured $P(\phi)$ is
most consistent with the analytical form in the range
$\kappa_{{\rm eff}}^{F}=2.32\pm0.08$, consistent with the relation
$\kappa=8(d_{f}-1)$. For L-PA domain walls measured from the fixed
end, we find $\kappa_{{\rm eff}}^{L}=2.85\pm0.10$.  We find
the same two BC-dependent values \cite{tobepub}
using another test: a comparison of the distribution for the
displacement between the DW endpoints with the form predicted using
dipolar SLE \cite{dipolar}, which describes the limit
$X/Y\rightarrow\infty$ for SLE curves that start at a given point $a$
and terminate on the upper boundary. Constraining the domain wall to
start at a given point (L-PA BCs) rather than {\em choosing} domain
walls that start at a point (i.e., conditioning on $a$) with F-PA BCs
changes the effective $\kappa$.  Domain walls with fixed endpoints are
not consistently described by $\mathrm{SLE}_\kappa$ over their entire
length.

The boundary conditions appear not to affect the fractal dimension, but
clearly do affect more subtle aspects of the geometry.
A similar result holds for the LERW:
absorbing and reflecting boundary conditions both are consistent
with $d_f=\frac{5}{4}$ \cite{BauerBernard}, but
$P(\phi)$ is well
fit by Eq.~(\ref{eq:angle}) with $\kappa=2$ for absorbing BCs, as
expected, in contrast with
$P(\phi)\approx \phi/\pi$ for reflecting BCs.

Note that the L-PA DW energy approaches a constant as
$L\rightarrow\infty$, in contrast with $E_{{\rm
DW}}\sim L^{\theta}$ for F-PA DWs.  This difference holds in general for
$\theta\le 0$, as can be seen by summing L-PA domain wall energies
defined over geometrically increasing scales connecting the
localized region to the large scale. Essentially, the L-PA
constraint gives an $O(1)$ correction to $E_{{\rm DW}}$ from the
cost of the single bond at the localized end. Apparently, the
optimization over successive scales distorts the curve from the form
expected from SLE, while optimization over a single global scale
gives results consistent with SLE.

\begin{figure}
\includegraphics[%
  width=1.0\columnwidth]{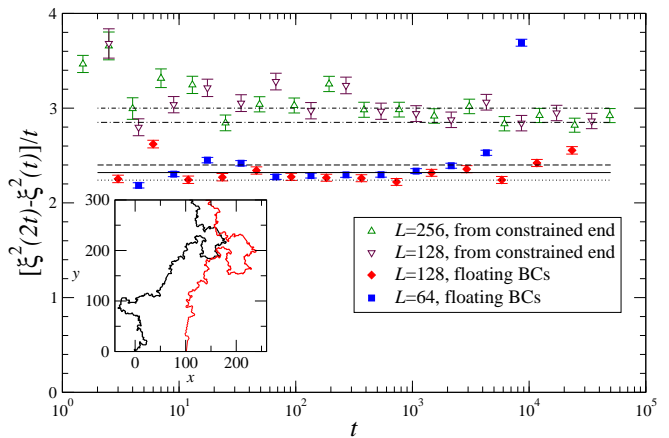}

\caption{\label{cap:keff}Plot of an effective diffusion constant
$\kappa_{{\rm eff}}=\overline{\xi^{2}(2t)-\xi^{2}(t)}/t$,
for $W\ge 4L$. Lines indicate
$\kappa=2.24$, 2.32, 2.40, 2.85, and 3.00.
The range $2.24<\kappa<2.40$ fits the data for curves with F-AP BCs, while
$2.85<\kappa<3.00$ describes the diffusion measured from a
constrained domain wall end.
Inset: Part of a sample conversion of a domain
wall in the 2D Ising spin glass to a sequence $\xi(t_{i})$, $i=1\ldots S$.
The left curve is the initial domain wall with $\xi(0)=0$,
while the red [lighter]
curve is the remainder after $500$ applications of
the dipolar map, giving $\xi(t_{500}\approx 7239.4)\approx 101.5$.}

\end{figure}

To more carefully inspect the correspondence with SLE, we have used a
discrete Loewner evolution to map the domain walls, represented by
sequences of points $z_{i}^{0}=x_{i}^{0}+iy_{i}^{0}$, $i=1\ldots S$,
in the complex half-plane, onto a real-valued sequence $\xi(t_{i})$
defined at discrete $t_{i}$ and studied the sample statistics and
correlations of the interpolated $\xi(t)$. For continuous
curves generated by SLE, the underlying function $\xi(t)$ is Brownian
motion with diffusion constant $\kappa$.
The sequence is
initialized by setting $t_{0}=0$ and $\xi(t_{0})=0$.  We then
recursively map the sequence $\{ z_{i}^{i-1},\ldots,z_{S}^{i-1}\}$ to
the transformed and shortened sequence $\{
z_{i+1}^{i},\ldots,z_{S}^{i}\}$ using the map appropriate for dipolar SLE
(first defining
$\Delta_i=\pi y_i^i/2Y$),
\begin{eqnarray}
t_{i} & = &
t_{i-1} - 2(Y/\pi)^2 \log[\cos(\Delta_i)]\ \ ;\ \ 
\xi(t_{i}) = x_{i}^{i-1}\label{eq:discretemap}\\
z_{j}^{i} & = &
\xi(t_{i})+
\frac{2L}{\pi}
\cosh^{-1}\left\{
           \cosh\left[\frac{\pi(z - \xi_i)}{2L}\right]
           /\cos(\Delta_i)\right\}.\nonumber
\end{eqnarray}
These maps are a sequence of slit maps that successively remove the
first point from the sequence (see Fig.~\ref{cap:keff}) and maintain
the hydrodynamic normalization used in SLE.

The simplest test for the diffusive property of $\xi(t)$ is to examine
its distribution at fixed times.  Our data for $L^{2}/5\agt t\agt50$
are consistent with a Gaussian distribution for $\xi(t)$ with
variance $\overline{\xi^{2}(t)}=\kappa_{{\rm eff}}t$
(Fig.~\ref{cap:keff}). We have confirmed that higher cumulants satisfy
$\overline{\xi^{2n}(t)}=(2n)!!\left(\overline{\xi^{2}(t)}\right)^{n}$
for $n=2,3,4$, within numerical error, for the same range of $t$.  For
computations of $\xi(t)$ that start from a free end of a domain wall
(F-PA boundary conditions {\em or} the free end of L-PA BCs), we find
$\kappa_{{\rm eff}}^{F}=2.24\pm0.08$, while for computations starting
from the localized end with L-PA BCs, we estimate
$\kappa_{{\rm eff}}^{L}\approx2.85\pm0.1$, consistent with our
estimates from $P(\phi,R)$. We note that $\xi^{2}(t)$ is also nearly
linear in time for paths on the MST, even though such paths are not
conformally invariant, but the coefficient is not consistent with the
fractal dimension (see \cite{Wilson} for MST winding angle results).

We have also tested the Markovian property for $\xi(t)$, i.e., that
the changes in $\xi(t)$ depend only on the current value of $\xi(t)$
and not on previous values.
We studied
the correlation function
$C_{d}(n)=\langle[\xi(t_{i+n+1})-\xi(t_{i+n})][\xi(t_{i+1})-\xi(t_{i})]\rangle$
at intermediate times; it
decays rapidly (by a factor of $\approx 100$
over the range $n=2$ to $n=8$) for both
the spin glass and for the LERW \cite{tobepub}.
Note that there must be short term correlations in $\xi(t)$
on the lattice, as there are
forbidden sequences of ``turns'' for the domain wall.

Given that the Ising spin glass DW passes several SLE tests one must
examine the domain Markov property (DMP) in a disordered system. Let
us call $P_{D}(\gamma_{ab})$ the probability that the DW happens to
coincide with the curve $\gamma_{ab}$ in a domain $D$ (where $a$,$b$
are two given boundary points). The DMP \cite{footnote0} states that
if one conditions this probability on a piece $\gamma_{ac}$ of the
curve, then the probability for the rest of the curve $\gamma_{cb}$ is
identical to the original probability on the cut domain
$D\setminus\gamma_{ac}$ conditioned on curves starting at $c$, i.e:
\begin{equation}
P_{D}(\gamma_{cb}|\gamma_{ac})=P_{D\setminus\gamma_{ac}}(\gamma_{cb}|c)
\label{dmp}
\end{equation}
Cutting the domain removes bonds that cross
the segment $\gamma_{ac}$.  In pure statistical
systems Eq.~\ref{dmp} is an identity, given proper
BCs.  One can easily check that the DMP holds
in a single realization of disorder.  However this property does not
survive disorder averaging (as conditioned probabilities are
ratio of probabilities) except for percolation $\mathrm{SLE}_{\kappa=6}$
(because of locality).

\begin{figure}
\includegraphics[%
  width=1.0\columnwidth]{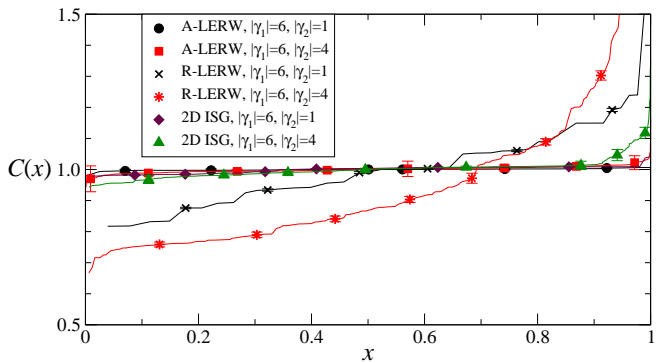}

\caption{\label{cap:markovtest} Plot of $C(x)$, cumulative probability
of ranked values for $r(\gamma_1,\gamma_2)$, as defined in the
text. Large deviations from $r=1$, as clearly seen for R-LERW,
indicate a failure of the domain
Markov property.
}
\end{figure}

To evaluate the deviations from the DMP we have computed numerically
the ratio of sums of the two probabilities in Eq. (\ref{dmp}). We generate
domain walls in both whole cylinders and in cylinders cut by all paths
$\gamma_1$ of a chosen path length $s_1$. We
sample at least $3\times 10^7$ disorder configurations to
estimate the ratio $r(\gamma_1,\gamma_2)=\sum_{\gamma_{bc} \supset
\gamma_2}P_{D\setminus\gamma_{1}}(\gamma_{bc}|c)/\sum_{\gamma_{bc}
\supset \gamma_2}P_{D}(\gamma_{bc}|\gamma_{1})$, with $\gamma_1$
starting at the lower boundary, $\gamma_2$ a subpath of $\gamma_{bc}$
of path length $s_2$, and $\gamma_1$ connecting to $\gamma_2$ at $c$.
If DMP holds strictly, $r(\gamma_1,\gamma_2)\equiv 1$.
We summarize our data in
Fig.\ \ref{cap:markovtest}, where we plot the cumulative probability
$C(x)=\sum_{\gamma1,\gamma_2|r(\gamma_1,\gamma_2)<x}P_{D}(\gamma_1,\gamma_2)$
that
$r$ is less than $x$.
We find that in small
samples with $L=8$, $r$ is statistically distinct from unity for larger $|\gamma_2|$.
The largest deviations are seen for
$\gamma_2$ near to and parallel to $\gamma_1$.
For comparison, we show results of the same analysis for the LERW with
both absorbing BCs (A-LERW),
where $r$ is unity within statistical error,
and for reflecting BCs (R-LERW), where $r$ clearly deviates from unity.
In ISG simulations, $C(x)$ is quite close to the curve for A-LERW.
Our data cannot rule out the possibility of DMP holding in the
continuum limit.

It is tempting to conjecture that the emergence of the DMP in the
continuum limit follows from the existence of ``principal''
minimizers, separated from each other on a scale $L$. These are the
basins of attraction for minimizing paths: if the start of a
DW is displaced from the minimizer's start by a scale $\ell<L$,
the DW merges with
principal minimizer within a distance of order $\ell$ \cite{middleton}. These
minimizers are a result of finding shortest paths in a graph with
negative weights (but no negative weight loops).  In particular, this
implies the same statistics for the L-PA and F-PA BCs on a long
cylinder. In a broad strip ($X\gg Y$), the differences between L-PA
and F-PA BCs must be related to the approach of the constrained path
to the principal minimizer over a sequence of scales.  Unlike local
minimizers, principal minimizers are independent of the direction in
which they are traversed.  We expect that bulk segments of the L-PA
curves are well described locally by $\mathrm{SLE}_\kappa$.  The
conditioning of paths used in defining the DMP may be related to the
properties of the minimizing paths \cite{footnoteproof}.  We also note
that the LERW with reflecting BCs passes the same set of tests of
conformal invariance as the L-PA 2DISG and fails the same set of tests
of $\mathrm{SLE}_\kappa$.

In conclusion, we have numerically sampled over geometric objects in a
system with disorder, domain walls in the 2D Ising spin
glass, and tested their statistical geometric properties. We find that
the domain walls pass to the left of a given point with probabilty
consistent with SLE, wind around long
cylinders in a manner consistent with conformal invariance, and 
that the sequences of conformal maps that generate DWs, i.e., Loewner
evolutions $\xi(t)$, give a diffusion constant
$\kappa=2.32\pm0.08$ in accord with a fractal dimension
$d_{f}=1.28\pm0.01$. We directly study the domain Markov
property: it fails in small systems, but we can not rule it out in
larger systems. This set of tests, whose utility is validated by
application to curves in LERW and MST, provides strong numerical
support for a description of spin-glass domain walls with
unconstrained endpoints by SLE, implying both conformal
invariance and a domain Markov property on long scales.
Domain walls starting from a
localized bond are not consistent with the simplest form of SLE,
though more complex conformally invariant descriptions, e.g., SLE with
drift such as $\mathrm{SLE}_{\kappa;\rho}$ should be investigated.

We thank M.~Biskup, M.~Bauer, J.~Cardy, C.~Newman, A.~Ludwig,
K.~Wiese, and T.~Witten and especially M.~Hastings for discussions
and the
authors of Ref.~\cite{amoruso} for sharing their unpublished
work. This work was supported in part by NSF grants DMR 0219292,
0606424, and ANR blan06-3-134462 and blan05-0099-01.  We thank the
KITP (NSF PHY99-07949) and the MPI-PKS for their hospitality.

{}

\begin{thebibliography}{100}
\bibitem{footnote1} 
This is consistent with $\kappa<4$ and zero double points on the
domain wall in the large scale limit.
\bibitem{footnote0}
Compared to standard SLE, the formulation of the
domain Markov property is modified for the case where the
starting point of the curve is not imposed.
Eq.\ \ref{dmp} is consistent with probabilities for curves
conditioned to start at a given point defined by dipolar SLE.
\bibitem{footnoteproof}
Conditioning upon $c$ reveals an interesting stability property of
minimizers to bond removal.
For a fixed bond configuration in $D$ with minimizer $\gamma_{ab}$,
we proved
that if the minimizer (DW) $\gamma'$ in the cut domain $D -
\gamma_{ac}$, where $\gamma_{ac}\subset \gamma_{ab}$,
starts at $c$, then (i) the ground states in $D$ and the cut
domain are the same and (ii) $\gamma'=\gamma_{cb}$.
\end{thebibliography}
\end{document}